\documentclass[prl,aps,twocolumn,showpacs,floatfix,reprint]{revtex4}
\usepackage{graphicx}
\usepackage{epsfig}
\usepackage{epstopdf}
\usepackage{amsmath}
\usepackage{amssymb}

\begin{document}
\title{Ionic Coulomb Blockade and Resonant Conduction in Biological Ion Channels}

\author{I.Kh.~Kaufman$^{1}$,
P.V.E.~McClintock$^1$,
R.S.~Eisenberg$^2$
}

\affiliation{$^1$Department of Physics, Lancaster University, Lancaster LA1 4YB, UK}
\email{p.v.e.mcclintock@lancaster.ac.uk}

\affiliation{$^2$Department of Molecular Biophysics and Physiology, Rush Medical College, 1750 West Harrison, Chicago, IL 60612, USA}

\date{\today}

\begin{abstract}

The conduction and selectivity of calcium/sodium ion channels are described in terms of ionic Coulomb blockade, a phenomenon based on charge discreteness and an electrostatic model of an ion channel. This novel approach provides a unified explanation of numerous observed and modelled conductance and selectivity phenomena, including the anomalous mole fraction effect and discrete conduction bands. Ionic Coulomb blockade and resonant conduction are similar to electronic Coulomb blockade and resonant tunnelling in quantum dots.  The model is equally applicable to other nanopores.

\end{abstract}



\maketitle

Biological ion channels are natural nanopores providing for the fast and highly selective permeation of physiologically important ions (e.g.\ Na$^+$, K$^+$ and Ca$^{2+}$) through cellular membranes. \cite{Hille:01, Ashcroft:06, Eisenberg:13a}. The conduction and selectivity of e.g.\ voltage-gated Ca$^{2+}$ \cite {Sather:03, Tang:14} and Na$^+$ channels \cite {Payandeh:11} are defined by the ions' movements and interactions inside a short, narrow selectivity filter (SF) lined with negatively-charged protein residues providing a net fixed charge $Q_f$. Permeation through the SF sometime involves the correlated motion of more than one ion  {\cite{Hodgkin:55, Armstrong:91, Roux:04, Kharkyanen:10}. It is known from mutation studies and simulations that $Q_f$  is a determinant of the channel's conductivity and valence selectivity \cite {Corry:05, Boda:08, Csanyi:12}. The discreteness of the ionic charge, and related electrostatic effects, play significant roles in ion channel conduction. It has recently been shown that nanopores can exhibit ionic Coulomb blockade (CB) \cite {Krems:13, Meyertholen:13}, a phenomenon equivalent to electronic CB in mesoscopic systems \cite {Alhassid:00, Beenakker:91, Pekola:13}.

An electrostatic theory describing ionic transport in water-filled periodically-charged nanopores has been proposed \cite{Zhang:05, Kamenev:06} treating the ions as a 1D Coulomb gas \cite {Kamenev:06}. It revealed the phenomenon of ion-exchange through low-barrier phase transitions as the ion concentration and fixed charge $Q_f$ \cite{Zhang:06} were varied. Comparable transitions in Brownian dynamics (BD) simulations of Ca$^{2+}$ channels result in discrete conduction and selectivity bands as functions of $Q_f$ \cite{Kaufman:13a,Kaufman:13b} consistent with earlier speculations \cite{Eisenberg:96} and explaining both the anomalous mole fraction effect (AMFE) \cite {Sather:03} and some of the puzzling mutation-induced transformations of selectivity in Ca$^{2+}$/Na$^+$ channels \cite{Heinemann:92, Tang:93, Shaya:11}. We have connected the bands' positions with sequential neutralisation and valence selectivity \cite {Kaufman:13b,Kaufman:13c}, but the physical origin of the bands and the statistical distribution of SF occupancy have remained unclear.

In this Letter, we will show that the permeation process in a simple model of the calcium/sodium ion channel is similar to mesoscopic transport in quantum dots: the simulated conduction bands and the experimentally observed valence selectivity phenomena in Ca$^{2+}$/Na$^+$ channels, including AMFE, can be well-described in terms of ionic CB conductance oscillations: the stop bands due to blockade are separated by resonant conduction bands, and the occupancy of the SF is governed by Fermi-Dirac (FD) statistics.

\begin{figure}[t]
\begin{center}
\includegraphics[width=1.0\linewidth]{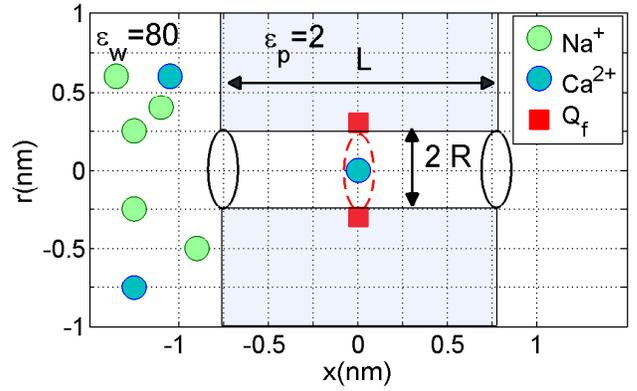}
\end{center}
\caption{(Color online)  Generic electrostatic model of the Ca$^{2+}$ or Na$^+$ channel. Ions inside the channel move in single file along its axis. For details, see text.} \label{fig:channel}
\end{figure}

In what follows, with SI units, $e$ is the electronic charge, $T$ the temperature, $z$ the ionic valence, $k_B$ Boltzmann's constant, and $\epsilon_0$ the vacuum permittivity.

We consider the generic electrostatic model of a Ca$^{2+}$/Na$^+$ ion channel shown in Fig.\ \ref{fig:channel}. The SF is described as an axisymmetric, water-filled, cylindrical pore of radius $R=0.3$nm ~and length $L=1.6$nm through a protein hub in the cellular membrane. A centrally-placed, uniform, rigid ring of negative charge $Q_f$ in the range $0 \leq |Q_f| \leq 7e$ is embedded in the wall at $R_Q=R$. The left-hand bath, modeling the extracellular space, contains non-zero concentrations of Ca$^{2+}$ and/or Na$^+$ ions. We take both the water and the protein to be homogeneous continua with relative permittivities $\epsilon_w=80$ and $\epsilon_p=2$, respectively, but describe the ions as discrete charges $q_i=ze$ within the framework of the implicit hydration model \cite{Parsegian:69,Laio:99, Roux:04, Zhang:05} moving in single file within the channel, with bulk values of diffusion coefficients $D_i$. We take no account of the negative counterions inside the SF, which will be few on account of repulsion by the negative $Q_f$. Such models are equally applicable to biological channels \cite {Boda:08,Corry:05} or to artificial nanopores \cite {Zhang:05, Krems:13, Zwolak:09, Corry:11, GarciaFandino:12}.

In the BD simulations, the coupled 3D axis-symmetrical Poisson electrostatic equation and 1D overdamped Langevin stochastic equation are solved numerically and self-consistently at each simulation step. The model obviously represents a considerable simplification of the actual electrostatics and dynamics of the ions and water molecules moving within the narrow SF \cite{Tieleman:01,Nelissen:07}. It was shown that, however, that $\epsilon$-electrostatics with bulk $\epsilon$ values works well for Ca$^{2+}$ and Na$^+$ channels ($R\ge0.3$nm) with the ions retaining their first hydration shells \cite {Laio:99, Corry:13, Tang:14}.  Details of the BD simulations and of the model, its range of validity and its limitations, have been presented and discussed elsewhere \cite{Kaufman:13a,Kaufman:13b}.

\begin{figure}[t]
\includegraphics[width=1.0\linewidth]{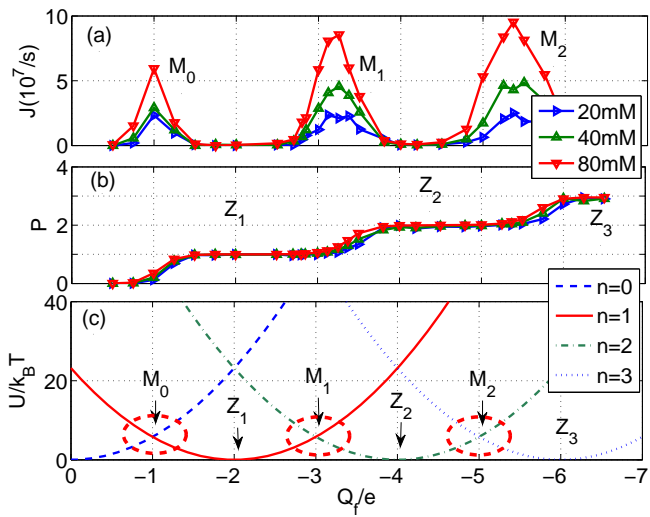}
\caption{(Color online)  Brownian dynamics simulations of multi-ion conduction and occupancy in a Ca$^{2+}$ channel model {\it vs} the effective fixed charge $Q_f$;
(a),(b) are reworked from \cite{Kaufman:13a}. (a) Plots of the Ca$^{2+}$ current $J$ for pure Ca$^{2+}$ baths of different concentration (20, 40 and 80mM as indicated). (b) The occupancy $P$. 
(c) The excess self-energy $U_{n,s}$ {\it vs} $Q_f$ for an empty channel $(n=0)$ and for channels with $n=1,2$ and 3 Ca$^{2+}$ ions inside.
The conduction bands $M_0$, $M_1$, $M_2$ and the blockade/neutralisation points $Z_1$, $Z_2$, $Z_3$ are discussed in the text.} \label{fig:ca_bands}
\end{figure}

The multi-ion conduction bands found in the BD simulations  \cite{Kaufman:13a,Kaufman:13b} are shown in Fig.\ \ref{fig:ca_bands}(a),(b) which plot the Ca$^{2+}$ current $J$ and channel occupancy $P$ for pure baths of different concentration. Fig.\ \ref{fig:ca_bands}(a) shows narrow conduction bands $M_0$, $M_1$, $M_2$  separated by stop-bands of almost zero-conductance centred on the blockade points $Z_1$, $Z_2$, $Z_3$. Fig.\ \ref{fig:ca_bands}(b) shows that the $M_n$ peaks  in $J$  correspond to transition regions in channel occupancy, where $P$ jumps from one integer value to the next, and that the stop-bands  correspond to saturated regions with integer $P=1,2,3...$. Band $M_0$ corresponds to single-ion conduction.  $M_1$ corresponds to the double-ion knock-on conduction, which is well-established for L--type Ca$^{2+}$ channels \cite{Armstrong:91,Sather:03}; and $M_2$ corresponds to triple-ion conduction which can be connected with Ryanodine receptor calcium channels \cite {Gillespie:08}. The bands can be considered as examples of self-organization in ion channels \cite {Giri:11, Tindjong:13b}

We can readily account for the  pattern of bands in terms of ionic CB, as shown in Fig.\ \ref {fig:ca_bands}(c). The discreteness of the ionic charge allows to us to introduce exclusive ``eigenstates" \{n\} of the channel with fixed integer numbers of ions inside its SF having total electrostatic energy $U_n$. The transition \{n\}$\rightarrow$ \{n+1\} corresponds to the entry of a new ion, whereas \{n\}$\rightarrow$ \{n-1\} corresponds to the escape of a trapped ion.
The statistics of ``eigenstates " is governed by an exclusive n-states definition \cite {Nelson:11} :
\begin{equation}
m=\{0,1,2,...\} \quad \sum_{m} \theta_m=1; \quad  P_c=\sum_{m} {m \theta_m},
\label {equ:set}
\end{equation}
where $\theta_m$ is the occupancy of the state \{m\} and $P_c$ is the average SF occupancy. In equilibrium $\theta_m$ is defined by the Boltzmann factor $\theta_m \propto \exp(-u_m)$ where $u_m=U_m/(k_B T)$. The exact distribution for $\theta_n$ and $P_c$ (which is FD) will be derived below.
The total energy $U_n$ for a channel in state \{n\} can be expressed as:
\begin {equation}
U_n=U_{n,s}+U_{n,attr}+U_{n,int}
\label {equ:un}
\end {equation}
where $U_{n,s}$ is the self-energy, $U_{n,attr}$ is the energy of attraction, and $U_{n,int}$ is the ions' mutual interaction energy.

The self-energy $U_{n,s}$ is defined on the assumption that both the ions and $Q_f$ are located within the central part of the SF, so that \cite {Zhang:05}:
\begin {equation}
U_{n,s}=\frac {1}{4 \pi \epsilon_0}
\frac {Q_n^2 L }{2 \epsilon_w R^2} ;
\quad Q_n=z e n +Q_f.
\label {equ:excess}
\end {equation}
Here, $Q_n$ represents the excess charge at the SF for the $n$ ions as function of $Q_f$.

To a first approximation we can omit  $U_{n,attr}$ and $U_{n,int}$, assume $U_n \approx U_{n,s}$, calculate $U_{n,s}$ as a function of $Q_f$ for $n=0,1,2,3$ and seek the conditions that minimize the self-energy.
Fig.\ \ref{fig:ca_bands}(c) plots $U_{n,s}$ as functions of $Q_f$. We note two kinds of low-energy singular points, marked as $M_n$ and $Z_n$.

The minima of $U_n$ (and the blockade regions) appear around the neutralisation points $Z_n=-z e n$ where $Q_n=0$ and the occupancy $P_c$ is saturated at an integer value \cite{Zhang:06, Kaufman:13c}. State \{n\} is separated from neighbouring \{n$\pm$1\} states by an impermeable barrier of $20k_BT$. The crossover points $M_n$ ($U_n=U_{n+1}$)  allow barrier-less $\{n\}\leftrightarrows\{n+1\}$ transitions; they correspond to the $P_c$ transition regions and to the conduction peaks in $J$ \cite {Kaufman:13c}. The $M_n$ points are separated from higher energy states by $\approx40k_BT$, which represents an impermeable barrier.

The positions of the singular points in Fig.\ \ref{fig:ca_bands}(a) can be written as:
\begin {equation}
\begin {split}
Z_n &=- z e  n \pm \delta Z_n, \quad \quad \quad \quad \text {Coulomb blockade}   \\
M_n &=- z e  (n+1/2) \pm \delta M_n \quad \text{Resonant conduction}
\end {split}
\label{equ:somm}
\end {equation}
where $\delta Z_n$, $\delta M_n$ are
possible corrections for affinity and for the ion-ion interaction, not accounted for here. Equation (\ref {equ:somm})  is exactly the same as its counterpart in electronic CB \cite {Alhassid:00}. We may therefore interpret the bands as arising from an alternation between ionic CB (stop-bands around the $Z_n$ points) and resonant conduction (the $M_n$ conduction bands) as $Q_f$ is increases. 

The positions of the peaks $M_n$ and stop-bands $Z_n$ in the simulations are consistent with an energetics analysis \cite{Kaufman:13b,Kaufman:13c}, supporting our interpretation of the observed conduction bands as ionic CB conductance oscillations \cite {Alhassid:00}; the small deviations in the precise positions of $M_n$ and $Z_n$ can reasonably be attributed to field leaks and simplifications inherent in the neutralisation approach.

Unlike its electronic counterpart, ionic CB is valence-dependent: the bands $M_n$ shift in proportion to $z$ due to (\ref {equ:somm}) and broaden/narrow in proportion to $z^2$ due (\ref {equ:excess}) \cite {Kaufman:13c}.
Fig.\ \ref {fig:CB}(a) plots the energy $U_n$ of an $n$-occupied channel against $n$ for the blockade/neutralisation points $Z_n$. For divalent Ca$^{2+}$ (full curve) $U_n$ has a sharp minimum for state \{n\}, separated from neighbouring \{n$\pm$1\} states by an impermeable barrier of $20k_BT$. This is strong blockade closely similar to electronic CB in quantum dots \cite {Alhassid:00, Beenakker:91}. The same plot for Na$^+$ ions (dashed curve) reveals a permeable $5k_BT$ barrier, that is weak ionic CB.

Fig.\ \ref {fig:CB}(b) show a similar plot for the resonant conduction points $M_n$. Again, it is strong ionic CB ($\Delta U\approx40k_BT$) for Ca$^{2+}$ ions and weak ionic CB ($\Delta U\approx10k_BT$) for Na$^{+}$ ions.

The ionic CB approach can be also applied to a mixed Ca$^{2+}$-Na$^+$ bath at the $Q_f=-3e$ point, which is $M_1$ for Ca$^{2+}$ and $Z_3$ for Na$^+$ ions. The combination  1Ca$^{2+}+$1Na$^+$ (with +3$e$ charge) inside the SF is neutralized  ($Q_n$=0) and so the channel is asymmetrically blocked due to ionic CB (Fig.\ \ref {fig:CB}(a)) -- strong for Ca$^{2+}$ and weak for Na$^+$. Similarly 3Na$^+$ (with +3$e$) also blocks the channel by weak ionic CB at the $Z_3$ point for Na$^+$ ions. Otherwise, a channel contents of 2Ca$^{2+}$ (+4e) gives $Q_n=+1e$ and resonant conduction (Fig.\ \ref {fig:CB}(b)). This asymmetric ionic CB provides an explanation of AMFE in Ca$^{2+}$ channels consistent with analysis in terms of the  potential energy landscape \cite {Kaufman:13b, Kaufman:13c}.

\begin{figure}[t]
\includegraphics[width=1.0\linewidth]{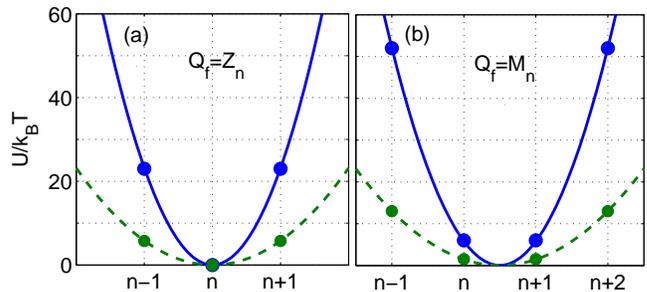}
\caption{(Color online) Valence-selective Coulomb blockade and resonant conduction. (a) Coulomb blockade. Plots of the self-energy $U_s$ for $Z_n$ points {\it vs} $n$, with Ca$^{2+}$ (blue), Na$^{+}$ (green), and quantization points (filled circles). (b) Resonant conduction. Plots of the self-energy $U_s$ plots for the $M_n$ points {\it vs} $n$ with $Ca^{2+}$ (blue), $Na^{+}$ (green), and quantization points (filled circles).
} \label{fig:CB}
\end{figure}

We are now in a position to derive the distribution of $P_c$ in the vicinity of the $M_n$ points and hence to calculate the shapes of $P_c(U)$ or $P_c(Q_f)$. For divalent Ca$^{2+}$, the separation in energy is large ($|U_{n-1}-U_n|\approx40k_BT$), so that general eigenstate definition (\ref {equ:set}) reduces to a discrete two-level exclusion principle:
\begin{equation}
m=\{n,n+1\}; \quad \theta_n+\theta_{n+1}=1; \quad  P_c=n+\theta_{n+1}.
\label {equ:Pauli}
\end{equation}
The electrostatic constraint (\ref{equ:Pauli}) plays the same role as the Pauli exclusion principle  plays in quantum mechanics \cite{Dirac:30, Kaniadakis:93, Liu:13a}. The standard derivation via a partition function, taking  account of (\ref {equ:Pauli}) leads \cite{Fowler:35} to  FD statistics for $\theta_{n+1}$ and an excess (fractional) occupancy $P^*_c=P_c \bmod 1$ :
\begin{equation}
P^*_c=
\frac{1}{1+P_b\text{exp}( u)}, \quad  u=u_{n+1}-u_n,
\label {equ:fd}
\end{equation}
where $P_b$ is some reference occupancy, related to bulk concentration. Note, that FD (\ref {equ:fd}) is equivalent to the Langmuir isoterm \cite {Fowler:35} (or Michaelis-Menten saturation). A similar Fermi function was obtained earlier \cite{Zhang:06} for the variation of $P_c$ with  concentration. Here we will assume $P_b$=1 for simplicity.

A self-consistent calculation of the conductance can be effected via the variance $\sigma$ of the occupancy $P$ due to thermal fluctuations.
The ability of an energy level to contribute to the current/conductance is proportional to $\sigma^2=dP/du$ via linear response theory and the Landauer approximation \cite {Alhassid:00, Beenakker:91}:
\begin{equation}
 J_{c}/J_{max}\propto dP/du=\text{cosh}^{-2}(u/2),
\label {equ:j0}
\end{equation}
where $J_{max}$ is the barrier-less diffusive current.

As an alternative, we can use the quasi-equilibrium (or nonequilibrium reaction rate \cite {Tindjong:12a}) approach with explicit solution of the Nernst-Planck equation for the triangular piece-wise linear approximation (i.e.\ Goldman-Hodgkin-Katz (GHK) solution) of the bell-shaped SF potential with account for FD occupancy (\ref {equ:fd}), yielding
\begin{equation}
J_{c}/J_{max}= \frac{u}{\sinh(u)}
 \label {equ:j1}
\end{equation}
which can be called the Fermi-GHK approximation and is identical to the corresponding approximation in the theory of CB \cite {Kulik:75, Beenakker:91}.

Fig.\ \ref{fig:FD_current} reveals a resonant conductivity as $U_c$ (or equivalently $Q_f$) is varied. The current $J_c$ in both the Landauer and Fermi-GHK approximations (Eqs.\ (\ref{equ:j0}) and (\ref{equ:j1})) exhibits a resonant peak coinciding with the maximum in the derivative of $P_c$, $dP/dU$. In practical terms, the difference between the two approximations is small. The form of this current is similar to that of the tunneling current in a quantum dot \cite {Alhassid:00}: an even, double-exponential function of $u$, reflecting the symmetry of escape and relaxation trajectories \cite {Luchinsky:97c}.
\begin{figure}[t]
\includegraphics[width=1.0\linewidth]{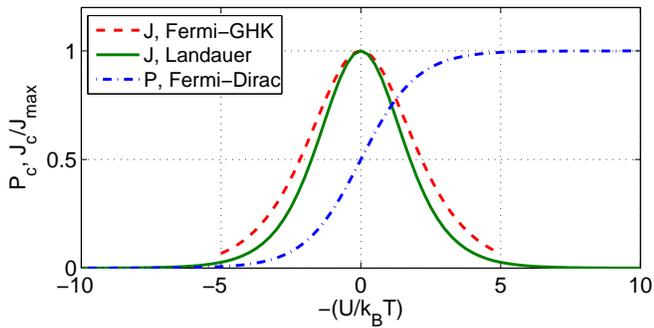}
\caption{(Color online) Calculations leading to resonant conduction as $U_c$ (or equivalently $Q_f$) is varied. The occupancy $P_c$ (blue, dash-dot) shows the FD transition from $P_c=0$ to $P_c=1$. The current calculated in either the Landauer $J_c\sim dP/dU$ (green, solid) or Fermi-GHK (red, dashed) approximation exhibits a resonant peak in the transitional region.} \label{fig:FD_current}
\end{figure}

For a quantitative fit of the theory to $P_c$, we calculate the effective (excess) well depth $U_c^*$ as:
\begin{equation}
U_c^* =k_B T \ln\frac {1-P^*_c}{P_c^*}
\label {equ:u_eff}
\end{equation}
The FD function (\ref{equ:fd}) predicts that $U^*_c$ should be linear in $U_c$ and also (due to the relatively narrow transition region) in $Q_f$, i.e.\ (\ref{equ:u_eff}) represents linearising coordinates for the FD equation (\ref{equ:fd}). Indeed, as shown in Fig.\ \ref{fig:FD_bands}(a), $U_c^*$ exhibits a piece-wise linear dependence on $Q_f$ with a high correlation coefficient $r\approx 0.98$, confirming that the $P^*_c$ transitions obey the FD function (\ref{equ:fd}) of $U_c$.

Fig.\ \ref{fig:FD_bands}(b) compares the ionic CB model with the the BD-simulated conduction bands $M_0$, $M_1$, $M_2$ \cite{Kaufman:13a}. Landauer and Fermi-GHK peaks are calulated using (\ref {equ:j0}) and (\ref {equ:j1}) respectively with the values of $u=U^*_c/k_bT$ taken from plot (a) and there are no adjustable parameters. The BD peak shapes and positions are described reasonably well by the model, extending the simpler fitting in \cite{Kaufman:13b, Kaufman:13c}. The discrepancies are attributable to our neglect of interaction effects.

\begin{figure}[t]
\includegraphics[width=1.0\linewidth]{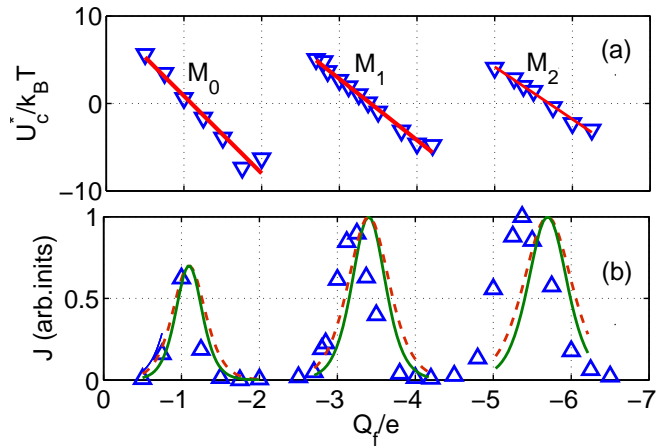}
\caption{(Color online) Comparisons of the ionic Coulomb blockade model with Brownian dynamics (BD) simulation results as $Q_f$ is varied. 
(a) The effective well depth $U_c^*$ (blue point-down triangles) fitted by Fermi-Dirac function (full red lines).  (b) The Landauer (green solid lines) and Fermi-GHK (red, dashed lines) peaks in $J$ compared  with BD simulation results (blue, point-up triangles).}
\label{fig:FD_bands}
\end{figure}

Although an ion moving inside a channel or nanopore is a classical system  described by Newtonian dynamics, it exhibits some quantum-like mesoscopic features including, in particular, ionic CB \cite {Krems:13, Meyertholen:13} at $Q_f=Z_n$, the FD distribution of $P_c$, and resonant barrier-less conduction at $Q_f=M_n$. We attribute such behavior to charge discreteness, confinement effects, and the fact that $U_n\approx k_BT$ while $|U_{n\pm1}-U_n|\gg k_BT$. It is interesting to note that 1D stochastic ionic dynamics inside the SF can be described by a Shr\"{o}dinger-like wave equation \cite {Kamenev:06}.

In conclusion, we have shown that Ca$^{2+}$ channel permeation is analogous to mesoscopic transport in quantum dots: the electrostatic exclusion principle leads to an FD distribution of channel occupancy; the stop-bands correspond to CB; the barrier-less conduction peaks are similar to those in resonant tunneling and can be described self-consistently in terms of the Landauer formula. The ionic CB model provides a good account of the experimental (AMFE and valence selectivity) and simulated (discrete multi-ion conduction and occupancy bands)  phenomena observed in model Ca$^{2+}$ channels. The results are should be applicable to other ion channels and to biomimetic nanopores with charged walls.

We are grateful to Will Gibby, Dmitrii Luchinsky and Rodrigue Tindjong for valuable discussions. The research was supported by the Engineering and Physical Sciences Research Council UK (grant No.\ EP/G070660/1).


\end{document}